\documentclass{emulateapj}
\usepackage{graphicx}
\usepackage{epsfig,psfrag}

\def\gsim{\;\rlap{\lower 2.5pt
 \hbox{$\sim$}}\raise 1.5pt\hbox{$>$}\;}
\def\lsim{\;\rlap{\lower 2.5pt
   \hbox{$\sim$}}\raise 1.5pt\hbox{$<$}\;}

\def\aap{A\&A}
\def\apj{ApJ }
\def\apjl{ApJL}
\def\apjs{ApJS}

\def\mnras{{MNRAS}}

\shorttitle{Cosmological Constraints from Cluster Velocities}
\shortauthors{Bhattacharya \& Kosowsky}

\begin{document}
\title{Cosmological Constraints from Galaxy Cluster Velocity Statistics}
\author{Suman Bhattacharya and Arthur Kosowsky}
\affil{ Department of Physics and Astronomy, University of
Pittsburgh, Pittsburgh, PA 15260} 
\email{sub5@pitt.edu, kosowsky@pitt.edu}

\begin{abstract}
Future microwave sky surveys will have the sensitivity to detect the kinematic Sunyaev-Zeldovich
signal from moving galaxy clusters, thus providing  a direct measurement of their line-of-sight
peculiar velocity. 
We show that cluster peculiar velocity statistics applied to foreseeable surveys will put significant 
constraints on fundamental cosmological parameters. We consider three statistical quantities
that can be constructed from a cluster peculiar velocity catalog: the probability density function, 
the mean pairwise streaming velocity, and the pairwise velocity dispersion. These quantities
are applied to an envisioned data set which measures line-of-sight cluster velocities with normal
errors of 100 km/s for all clusters with masses larger than $10^{14}$ solar masses over
a sky area of up to 5000 square degrees. A simple Fisher matrix analysis of this survey shows that
the normalization of the matter power spectrum and the dark energy equation of state can be
constrained to better than 10 percent, and the
Hubble constant and the primordial power spectrum index can be constrained to  a few percent, independent of any other cosmological observations. We also find that  the current constraint on the power spectrum normalization can be improved by more than a factor of two using data from
a 400 square degree survey and WMAP third-year priors. We also show how the constraints on cosmological parameters changes if cluster velocities are measured with normal errors of 300 km/s.
\end{abstract}


\section{Introduction}

One of the great triumphs of cosmology has been a precise determination of the parameters
defining the standard cosmological model, primarily via the cosmic microwave background fluctuations
combined with the large-scale distribution of galaxies and the distance-redshift relation of
distant supernovae. Future high-precision probes of
cosmological parameters using different sources of data provide the possibility of tighter constraints on the cosmological parameters, along with strong consistency checks of the cosmological model. Here
we consider one such potential probe, the peculiar velocities of galaxy clusters.  Upcoming microwave sky surveys like Planck, ACT \cite{kosowsky06,fowler05,kosowsky03} and SPT \cite{ruhl05} will detect thousands
of galaxy clusters via their thermal Sunyaev-Zeldovich distortions of the microwave background
frequency spectrum \cite{SZ80}. 
With sufficient sensitivity, angular resolution, and control of systematic errors,
microwave surveys also have the potential to measure directly the line-of-sight peculiar velocity of clusters via their kinematic Sunyaev-Zeldovich signature, a near-blackbody distortion of the microwave
background proportional to the optical depth and velocity of the cluster. Since the SZ effect is
effectively independent of cluster redshift, such a survey can probe the cosmic velocity field
over a cosmological volume of the universe.

In practice, future data sets will give sky positions, redshifts, and the line-of-sight peculiar velocity
component for massive galaxy clusters above a certain SZ flux threshold, which will
roughly correlate with a cluster mass limit. Massive clusters are the largest gravitationally bound
objects in the universe, and their distribution and velocities probe the mildly nonlinear regime of
structure formation. For given initial perturbations in the early universe, peculiar velocities at
recent epochs depend on the expansion history of the universe. Resulting constraints on
cosmological parameters will complement those from microwave background measurements
(e.g.\ \cite{WMAP3}) and from supernova searches (e.g.\ \cite{SNLS1}). Cluster number counts
from their thermal SZ signature 
also show this complementarity, but they are prone to bias from systematic errors in
cluster mass determination (Francis, Bean \& Kosowsky 2005). Cluster peculiar velocities are much less sensitive to
this systematic error, and so may provide a more promising route to cosmological constraints,
even though the kinematic SZ signal is much smaller than the thermal SZ signal.

Other recent studies in this area \cite{peel06,penn06,peel02} laid out the theoretical framework for
studying the correlation function of the cosmic velocity field through the cluster kSZ effect while DeDeo, Spergel \& Trac 2005 studied cross-correlations of kSZ signal with galaxy counts. This paper explores how well various velocity statistics for a galaxy cluster velocity survey can constrain cosmology.

\section{Theory}

We will consider three different velocity statistics: the probability density function of the line-of-sight component of peculiar velocities $f(v)$, and the mean pairwise streaming velocity $v_{ij}(r)$ and  pairwise velocity dispersion $\sigma_{ij}(r)$ for two clusters separated by a distance $r$. These are
formed by averaging over all clusters or cluster pairs in a given peculiar velocity survey. In the
halo model, these quantities can be written as the sum of the contribution from one-halo and two-halo terms. 
Since we are interested in very massive clusters, the one-halo term can be neglected. When compared with the VIRGO numerical simulation of structure formation \citep{virgo,virgo1}, the two-halo term approximates
the probability density function fairly well, and for $v_{ij}(r)$ and $\sigma_{ij}(r)$ it agrees well for separations $r$ larger than 20 Mpc and 30 Mpc, respectively.

Analytic approximations for all three of these statistics in dark matter-dominated cosmological models
give their dependence on parameters describing the background cosmology.
The probability density function of the line-of-sight peculiar velocity component at some redshift $z$
is given by 
\cite{sheth02}
\begin{equation}
f(v,z)=\frac{\int  dm\,mn(m|\delta)p(v|m,\delta)}{\int dm\,mn(m|\delta)}
\end{equation}
where $n(m|\delta) dm$ is the number density of halos that have mass between $m$ and $m+dm$ in
a region with overdensity $\delta$,  and $p(v|m,\delta)$ is the probability that halos of mass $m$ in a local overdensity $\delta$ move with line-of-sight velocity $v$; the dependence of these
quantities on redshift is left implicit. Assuming Gaussian initial conditions,
we can write $n(m|\delta)$ in terms of the total number density $n(m)$ and a large-scale bias factor 
$b(m)$, 
\begin{equation}
n(m|\delta) \approx \left[1+b(m)\delta\right]n(m).
\end{equation}  
Define moments of the initial mass distribution with power spectrum $P(k)$ by \cite{bardeen86} 
\begin{equation}
\sigma_j^2(m)\equiv \frac{1}{2\pi^2}\int_0^{\infty}{dk ( k^{2+2j})P(k)W^2(kR(m))}
\end{equation}
when smoothed on the scale $R(m)= (3m/4\pi\rho_0)$ with the top-hat filter  
$W(x)=3[\sin(x)-x\cos(x)]/x^3$, and $\rho_0$ the present mean matter density. 
Then the bias factor can be written as \citep{sheth04} 
\begin{equation}
b(m)=1+\frac{\delta_{\rm crit}^2-\sigma_0^2(m)}{\sigma_0^2(m)\delta_{\rm crit}D_a}.
\end{equation}
where $D_a$ is the linear growth factor at scale factor $a$, normalized to $1$ today, and the
critical overdensity $\delta_{\rm crit}\approx 1.686$. 
For the number density of halos $n(m)$, we use the Jenkins mass function \citep {jenkins01}
\begin{eqnarray}
\lefteqn{\frac{dn}{dm}(m,z)= 0.315\frac{\rho_0}{m^2}\frac{d \ln \sigma_0(m)}{d \ln m}} \nonumber \\
 &\times& \exp\left[-\left|0.61-\ln(\sigma_0(m) D_a)\right|^{3.8}\right].
\end{eqnarray}
This mass function is a fit to numerical simulations of cold dark matter gravitational clustering. 

Along with $n(m|\delta)$ just defined, we also need $p(v|m,\delta)$, which we assume to
be gaussian \citep{sheth02},
\begin{equation} 
p(v/m,\delta)= \frac{\exp^{-[3v/\sigma_v(m)]^2}/2}{\sqrt{2\pi \sigma_v^2(m)/3}} 
\end{equation}
with  the three-dimensional velocity dispersion smoothed over a length scale $R(m)$ given by \\\cite{hamana03}
\begin{eqnarray}
\sigma_v(m,a)&=& \left[1+\delta(R_{\rm local})\right]^{2\mu(R_{\rm local})}aH(a)D_a \frac{d \ln D_a}{d \ln a}\nonumber\\
 &\times& \left(1-\frac{\sigma_0^4(m)}{\sigma^2_1(m)\sigma^2_{-1}(m)}\right)^{1/2}\sigma_{-1}(m)
\end{eqnarray}
Here $H(a)$ is the Hubble parameter as a function of scale factor, $R_{\rm local}$ is a smoothing scale with which the local background density $\delta$ is defined, and $\mu(R_{\rm{local}})\equiv 0.6 \sigma^2_0(R_{\rm{local}})/\sigma^2_0(10\,{\rm Mpc}/{\rm h})$ \\\cite{sheth02}. Following \cite{hamana03}, $R_{\rm local}$ is obtained empirically using N-body simulations via the condition 
$\sigma_0(R_{\rm{local}})=0.5(1+z)^{-0.5}$. We have now completely specified $f(v,z)$ in terms of quantities related to the background cosmology.

The mean relative peculiar velocity $v_{12}(r)$ for all pairs of halos at comoving separation $r$  can be related to the linear two-point correlation function $\xi^{\rm lin}(r,a)$ for dark matter using large-scale bias and the pair conservation equation \citep{sheth04}:
\begin{equation}
v_{12}(r)=-\frac{2}{3}aH(a)\frac{d \ln D_a}{d \ln a}\frac{rb_{\rm halo}\bar{\xi}^{\rm lin}(r,a)}{1+b_{\rm halo}^2\xi^{\rm lin}(r,a)}
\end{equation}
where $b_{\rm halo}\equiv \int dm\,mn(m)b(m)/\int dm\,n(m)$ is the average halo bias factor
and $\bar{\xi}^{\rm lin}(r,a)$  is the two-point correlation function averaged over a ball of radius $r$.
Under linear evolution, the correlation function simply scales as
\begin{equation} 
\xi^{\rm lin}(r,a)=D_a^2 \xi^{\rm lin}(r,a_0).
\end{equation}

Similarly, a model for  the pairwise velocity dispersion $\sigma_{12}(r)$, in the large-scale limit and neglecting velocity correlations, is given by \citep{sheth01}
\begin{equation}
\sigma^2_{ij}(r)=2\frac{ \left\langle\sigma_v^2(R,a)\right\rangle +\xi^{\rm lin}(r,a)\left\langle b\sigma_v^2(R,a)\right\rangle}{1+b_{\rm halo}^2\xi^{\rm lin}(r,a)}
\end{equation}
where $\left\langle\cdots\right\rangle$ represents an ensemble average over all 
halos lying within a given mass range.
 Note the above expressions for  $v_{12}(r)$ and $\sigma^2_{ij}(r)$ are for three dimensional velocity distributions. \cite{ferreira99} give one technique for constructing an estimator of $v_{12}(r)$ using only line-of-sight velocities; estimators for both $v_{12}(r)$ and $\sigma^2_{ij}(r)$ will be presented elsewhere (Bhattacharya et al. in prep.).

These analytic approximations to the cluster velocity probability distribution function, the
cluster pairwise mean relative velocity, and the pairwise mean velocity dispersion explicitly
display their cosmological dependences through the Hubble parameter $H(a)$, the growth
factor $D_a$, the mean mass density $\rho_0$, and the power spectrum $P(k)$. Note that they depend
on the background cosmology in significantly different ways.

\section {Estimating Parameter Constraints}

To understand the dependence of the distribution of velocities on the underlying cosmology, we perform a
simple Fisher matrix analysis for the three velocity statistics described in the previous Section. Assuming a flat Universe,the set of cosmological parameters $\bf p$ on which the velocity field depends are the normalization of the matter power spectrum, conventionally parameterized by $\sigma_8$, the power law index of the primordial
power spectrum $n_S$, the total matter density $\Omega_m$, the dark energy equation of state $w$ and the Hubble parameter $h$. 
We consider a  fiducial model with $\sigma_8=0.9$, $n_S=1$, $\Omega_m=0.3$, $w=-1$, $h=0.7$. Upcoming data sets will provide cluster number counts and line-of-sight velocities as a function of redshift; we assume
here that all clusters with $M>10^{14}M_\odot$ are detected, that redshifts are measured with
zero error for $z < 1.5$ and that peculiar velocity is measured with a standard error of 100 km/sec, 
the limit from internal cluster motions (Nagai, Kravtsov \& Kosowsky 2003). This is likely a realistic mass limit
for high-sensitivity microwave surveys \cite{sehgal07}. The velocity error represents
the fundamental limit from intrinsic cluster properties; measurements to this precision will
likely require measurements of cluster temperature (Sehgal, Kosowsky \& Holder 2005) and detailed understanding
of the point sources and other signals which can contaminate the kinematic SZ signal
\cite{coble06,holder04}. However \cite{diaferio05} have extracted velocities from cluster simulations using multiple microwave and X-ray bands finding a robust lower limit of 200 km/s. We also show how constraints on different parameters degrades for a larger standard error of 300 km/s.   
 Spectroscopic redshifts will be challenging to obtain for clusters beyond
$z=1.5$, so we consider clusters only to this distance. We approximate the cosmic variance error for each observables to be proportional to $(r/L)^3$, where $r$ is the separation between two halos and $L$ is the size of a particular survey. A detail study of cosmic variance will be discussed elsewhere.

\begin{table}
\begin{center}
\begin{tabular} {l|l|l|l} 
\hline 
Parameter & 5000 deg$^2$ & 1000 deg$^2$ & 400 deg$^2$ \\ \hline 
$\sigma_{8}$ [5.6\%] & 7.2\% \{2.1\%\} & 16.0\% \{2.4\%\} & 25.6\% \{2.4\%\}\\
$n_S$ [1.5\%] & 2.2\% \{1.2\%\} & 4.9\% \{1.4\%\} & 7.7\% \{1.4\%\}\\ 
$\Omega_{\rm m}$ [6.7\%] & 7.9\% \{4.7\%\}& 17.9\% \{6.0\%\}& 28.2\% \{6.3\%\}\\
w [10\%] & 9.2\% \{6.0\%\}& 20.5\% \{8.6\%\}& 31.6\% \{9.3\%\}\\ 
$h$ [4.3\%] & 18.9\% \{4.0\%\} & 42.4\% \{4.1\%\} & 67.0\% \{4.0\%\}\\
\end{tabular}
\end{center}
\caption {\rm $1\sigma$ errors in cosmological parameters from $f(v)$ for clusters with masses
$M>10^{14}M_\odot$, measured line-of-sight velocity error $\sigma_v=100$ km/s, and $z<1.5$, for three survey areas. Percent errors are given in terms of the fiducial values. Errors in braces include a Gaussian prior from current measurements, given in square brackets.}
\end{table}

\begin{table}
\begin{center}
\begin{tabular}{l|l|l|l}
\hline 
Parameter &5000 deg$^2$& 1000 deg$^2$& 400 deg$^2$\\ \hline 
$\sigma_{8}$ [5.6\%] & 34.8\% \{3.5\%\} & 174.0\% \{3.9\%\} & 435.0\% \{4.7\%\} \\ 
$n_S$ [1.5\%] & 1.87\% \{0.8\%\} & 9.4\% \{1.3\%\} & 23.5\% \{1.4\%\} \\ 
$\Omega_{\rm m}$ [6.7\%] & 18.2\% \{4.6\%\} & 92.0\% \{6.3\%\} & 230.7\% \{6.5\%\} \\ 
w [10\%] & 16.7\% \{7.6\%\} & 84.3\% \{9.64\%\} & 211.0\% \{9.8\%\}\\ 
$h$ [4.3\%] & 42.4\% \{3.6\%\} & 212.0\% \{3.75\%\} & 531.0\% \{4.0\%\}\\
\end{tabular}
\end{center}
\caption[Table 2]{\rm Same as in Table 1, but for $v_{ij}(r)$.}
\end{table}

\begin{table}
\begin{center}
\begin{tabular}{l|l|l|l}
\hline
Parameter &5000 deg$^2$& 1000 deg$^2$& 400 deg$^2$ \\ \hline 
$\sigma_{8}$ [5.6\%] & 20.2\% \{3.0\%\}& 102.0\% \{3.5\%\} & 255.0\% \{3.7\%\}\\ 
$n_S$ [1.5\%] & 14.8\% \{1.4\%\} & 74.7\% \{1.4\%\} & 187.0\% \{1.4\%\}\\ 
$\Omega_{\rm m}$ [6.7\%] & 14.1\% \{5.6\%\} & 71.2\% \{6.2\%\} & 176.0\% \{6.3\%\}\\ 
w [10\%] & 8.3\% \{5.6\%\} & 41.6\% \{9.4\%\} & 104.0\% \{9.8\%\} \\ 
$h$ [4.3\%] & 3.2\% \{2.3\%\} & 16.2\% \{3.8\%\} & 40.4\% \{3.9\%\}\\
\end{tabular}
\end{center}
\caption[Table 3]{\rm Same as in Table 1, but for $\sigma_{ij}(r)$.}
\end{table}

\begin{table}
\begin{center}
\begin{tabular}{l|l|l|l}
\hline
Parameter & $f(v)$ & $v_{ij}(r)$ & $\sigma_{ij}(r)$ \\ \hline 
$\sigma_{8}$ [5.6\%] & 7.3\% \{2.2\%\} & 61\%\{5.0\%\} & 36.8\%\{3.7\%\} \\
$n_S$ [1.5\%] & 2.9\%\{1.4\%\} & 3.3\%\{1.1\%\} & 28.1\%\{1.4\%\} \\ 
$\Omega_{\rm m}$ [6.7\%] & 10.5\%\{6.0\%\} & 25\%\{5.5\%\} & 21.1\%\{6.1\%\} \\ 
w [10\%] & 12.3\%\{7.0\%\} & 24.7\%\{8.3\%\} & 11.1\%\{7.8\%\} \\ 
$h$ [4.3\%] & 23.0\%\{4.0\%\} & 56.2\%\{4.0\%\} & 5.6\%\{2.9\%\} \\
\end{tabular}
\end{center}
\caption[Table 4]{\rm $1\sigma$ errors in cosmological parameters obtained from the three velocity statistics for 5000 deg$^2$ with measured line-of-sight velocity error $\sigma_v=300$ km/s.}
\end{table}

The measured observables are the value of $f(v)$ in redshift and velocity bins, and the values of
$v_{ij}(r)$ and $\sigma_{ij}(r)$ in redshift and distance bins; the variance in each bin depends
on Poisson errors, cosmic variance, and velocity errors. The  redshift bins $\Delta z=0.2$ are chosen to be narrow enough to capture statistically significant redshift evolution. Then assuming a normal error distribution
in each bin, the Fisher information matrix for each statistic is simply \cite{jungman96}
\begin{equation}
F_{\alpha\beta}= \sum_{k,l}\frac{\partial \phi(x_k,z_l)}{\partial p_\alpha}\frac{1}{\sigma^2_{\phi}}\frac{\partial \phi(x_k,z_l)}{\partial p_\beta}
\end{equation}
where the values $z_l$ define
the redshift bins, the values $x_k$ define the bins in either velocity (for $f$)
or distance (for $v_{ij}$ and $\sigma_{ij}$), $\phi$ is either $f(v,z)$, $v_{ij}(r,z)$ or $\sigma_{ij}(r,z)$,
$\sigma^2_\phi$ is the variance of the function $\phi$ in each bin,  and the partial derivatives are evaluated
for the fiducial values of the cosmological parameters. The inverse of the Fisher matrix
then gives an estimate for the variances of each cosmological parameter, marginalized over
the values of the other parameters, along its diagonal, and
correlations between the parameters in its non-diagonal elements. 

We have looked at parameter constraints for cluster velocity surveys of three different
sizes: sky areas of 400, 1000, and 5000 square degrees. We calculate the
parameter errors for each of the three velocity statistics individually, and
also when imposing prior constraints on each parameter taken from the WMAP
3-year data \citep{WMAP3} plus current large-scale structure and supernova measurements
\citep{SNLS1}. Results are summarized in Tables 1,2,and 3 for each of the three velocity statistics. Table 4 shows the constraints for a 5000 deg$^2$ survey when the velocity error is increased to 300 km/s. Comparing the results for 300 km/s with those for 100 km/s, we find that for $f(v)$ the errors on parameters changes by 3-4 \% while for $v_{ij}(r)$ and $\sigma_{ij}(r)$ this degradation is more pronounced. However note that most of the constraints on cosmological parameters comes from $f(v)$, so the overall errors on cosmological parameters degrades by 3-4\%.

It is clear from these results that a number of cosmological parameters can be
well constrained by large cluster peculiar velocity surveys. In particular, both
$f(v)$ and $v_{ij}$ can provide strong constraints on $n_S$; for a survey
covering 5000 square degrees, the constraints of 2.2\% and 1.9\% are close to
the current precision from microwave background measurements. Note that these
constraints on $n_S$ are at a smaller physical scale than the corresponding constraints 
from primary microwave anisotropies, and a mismatch in these two high-precision
measurements indicates a departure from a pure power law of the primordial 
density perturbations. The Hubble parameter
can be measured using $\sigma_{ij}$ to 3.2\%, better than current constraints.
All three statistics give constraints on $w$ to around 10\%, competitive
with current measurements. Finally, $f(v)$ gives a measurement of $\sigma_8$ to
7.2\%, competitive with current limits. These limits are for each parameter,
marginalized over the others. 


%

\section{Conclusion and Prospects}

Galaxy cluster peculiar velocities, measured directly via the kinematic Sunyaev-Zeldovich
effect, represent a new route to precision cosmological constraints. We have shown here
that if cluster line-of-sight velocities can be measured with errors of 100 km/sec, the
resulting constraints on several cosmological parameters will be comparable with
all current techniques. Such measurements could be important for their constraints
on particular parameters, but are likely more valuable as consistency checks on
the standard cosmological model. Multiple measurements of
each cosmological parameter using independent methods provides our only
way to determine whether our universe is actually described by
the simple models spanned by the standard
cosmological parameter space. The awkward appearance of dark energy on
the cosmological stage makes these cross-checks all the more imperative. 

A potentially even more important advantage to peculiar
velocities is the control of systematic errors. It  has been appreciated
for some time that the comoving number density of clusters above a certain
mass limit as a function of redshift depends sensitively on the underlying
cosmology: a small increase in the cosmological growth factor leads to
a large increase in the number of clusters (Viana \& Liddle 1996; Eke et al 1996; Barbosa et al 1996; Bahcall, Fan \& Cen 1997).
Blind Sunyaev-Zeldovich surveys will detect
all clusters above a certain SZ-distortion threshold in a given
direction of the sky, so their signal also depends sensitively on cosmological
parameters. But the cluster SZ
selection function is not equivalent to a simple mass cutoff, and
a small systematic error in understanding this selection function can result
in substantial systematic errors in cosmological constraints (Francis, Bean \& Kosowsky 2005; Holder et al 2001).
In contrast, the velocity statistics considered here have little dependence
on the cluster selection function, as we will show explicitly elsewhere; 20\% variations in the mass selection function give only a few percent
change in the velocity statistics.

The observational prerequisite for measuring galaxy cluster velocities are microwave
observations with multiple frequencies, arcminute angular resolution, and sensitivity
of a few $\mu$K per arcminute pixel. The Atacama Cosmology Telescope (ACT)
will nominally meet these requirements \cite{kosowsky06}; it is scheduled for deployment in Chile
in the first part of 2007 (although it will not commence three-frequency observations
until 2008). The South Pole Telescope (SPT) also could make sufficiently sensitive maps,
although its nominal initial survey will be substantially wider and shallower than
ACT \cite{ruhl05}. Within the next few years, we will have the raw experimental capability to
construct cluster peculiar velocity catalogs. The extent to which this program will
yield the kinds of cosmological constraints discussed here will then depend on our
ability to extract the small velocity signal from the mix of other larger signals at
multiple wavebands, including the clusters' thermal SZ distortion, infrared and
radio point sources (Coppin et al 2006; Knox, Holder \& Church 2004), and gravitational lensing by the cluster \cite{holder04}. The remarkable strides in observational cosmology over the past ten years suggest that optimism
is justified. 

\section*{Acknowledgments}
We acknowledge many useful discussions with Ravi Sheth and Ryan Scranton. We also acknowledge Toby Marriage, David Spergel, Raul Jimenez and Joseph Fowler for providing many helpful suggestions about an earlier draft. This work has
been supported by NSF grant AST-0408698 to the ACT project, and
by NSF grant AST-0546035.

\end{document}